\documentclass[a4paper,11pt]{article}
\usepackage{pos}
\usepackage{subcaption}
\usepackage{enumerate}
\usepackage{aas_macros}
\usepackage{url}

\title{Tidal Disruption Events and High-Energy Neutrinos}
\ShortTitle{TDEs and Neutrinos}

\author*[a, b]{Robert Stein}

\affiliation[a]{Deutsches Elektronen-Synchrotron DESY: Zeuthen, Germany}
\affiliation[b]{Institut f\"ur Physik, Humboldt-Universit\"at zu Berlin: Berlin, Germany}

\emailAdd{robert.stein@desy.de}

\abstract{Tidal Disruption Events (TDEs) occur when stars pass close to supermassive black holes, and have long been predicted to emit cosmic rays and neutrinos. Recently the TDE AT2109dsg was identified in spatial and temporal coincidence with high-energy neutrino IC191001A as part of the Zwicky Transient Facility (ZTF) neutrino follow-up program, providing the first direct observational evidence supporting these objects as multi-messenger sources. In these proceedings, I will place the recent results of our ZTF neutrino follow-up program into the broader context of developments in TDE and neutrino astronomy.}

\FullConference{37$^{\rm{th}}$ International Cosmic Ray Conference (ICRC 2021)\\
		July 12th -- 23rd, 2021\\
		Online -- Berlin, Germany}

\begin{document}
\maketitle

\section{Tidal Disruption Events}

A Tidal Disruption Event (TDE) occurs when a star passes sufficiently close to a supermassive black hole (SMBH), such that the tidal forces exceed the self-gravity holding the star together \cite{rees_tde_88}. The star then disintegrates and is partially accreted onto the SMBH, generating an electromagnetic flare that can then be detected on Earth \cite{gezari_21}. TDEs have also been observed to launch relativistic jets \cite{swift_j1644_11} or outflows \cite{van_velzen_16, alexander_16, radio_tde_summary}, but these signatures are not ubiquitous. Indeed, it is an area of active research to understand when such outflows or jets are launched \cite{radio_tde_summary}.

\begin{figure}[!ht]
	\centering \includegraphics[width=0.45\textwidth]{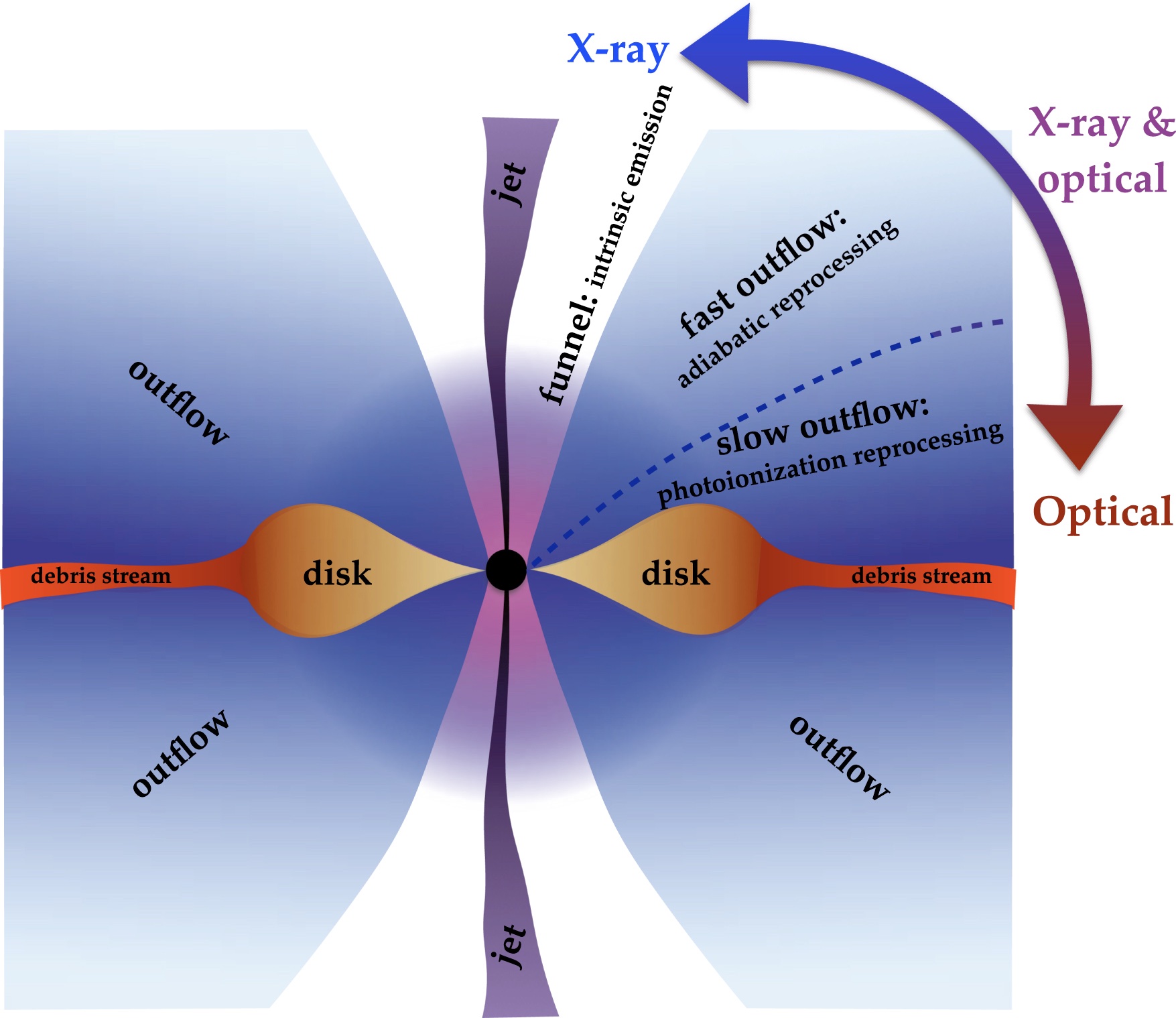}
	\caption{The unified model of TDEs, from \cite{dai_18}.}
	\label{fig:tde_dai}
\end{figure}

In order to understand the properties of TDEs, candidates must first be identified by telescopes, so one might reasonably ask `\emph{what does a TDE look like?}'. It turns out that the answer is far from straightforward. Models for electromagnetic radiation from TDEs tend to fall into two broad classes, depending on whether the stellar debris is thought to rapidly circularise or instead remains highly eccentric \cite{roth_20}. For the case of `rapid circularisation', the electromagnetic radiation can be explained within the framework of a `unified TDE model' \cite{dai_18}, as shown in Figure \ref{fig:tde_dai}, somewhat akin to the blazar unification models \cite{95_agn_unification}. There would be a central SMBH with a rapidly-formed accretion disk, and a stellar debris stream, for which a substantial fraction of radiation will be reprocessed. A TDE viewed side-on might only exhibit reprocessed UV/Optical emission, while a TDE viewed face-on could exhibit bright X-ray emission, for example from a jet. For TDEs viewed at intermediate viewing angles, one would expect a mixture of these scenarios. Alternatively, the stellar debris might not rapidly form an accretion disk. Instead, the electromagnetic emission could arise from shocks in the debris \cite{roth_20}. 

While TDEs were first predicted in the late 1980s \cite{rees_tde_88}, actually detecting them in great numbers has long proved challenging. It is interesting to consider what would have been written about TDEs at the previous 36th ICRC in 2019. At that time, there were just $\sim$2 dozen `TDE candidates' reported in the literature, detected across a range of observatories and wavelengths. The vast majority had relatively sparse lightcurves and little multi-band coverage, while only a couple had been detected pre-peak. However, a then-recent publication from the Zwicky Transient Facility (ZTF) survey was confidently predicting that a step change in TDE discovery rates lay just around the corner, with an imminent doubling of TDE sample size \cite{ztf_19_science}.

\begin{figure}[!ht]
	\centering \includegraphics[width=0.55\textwidth]{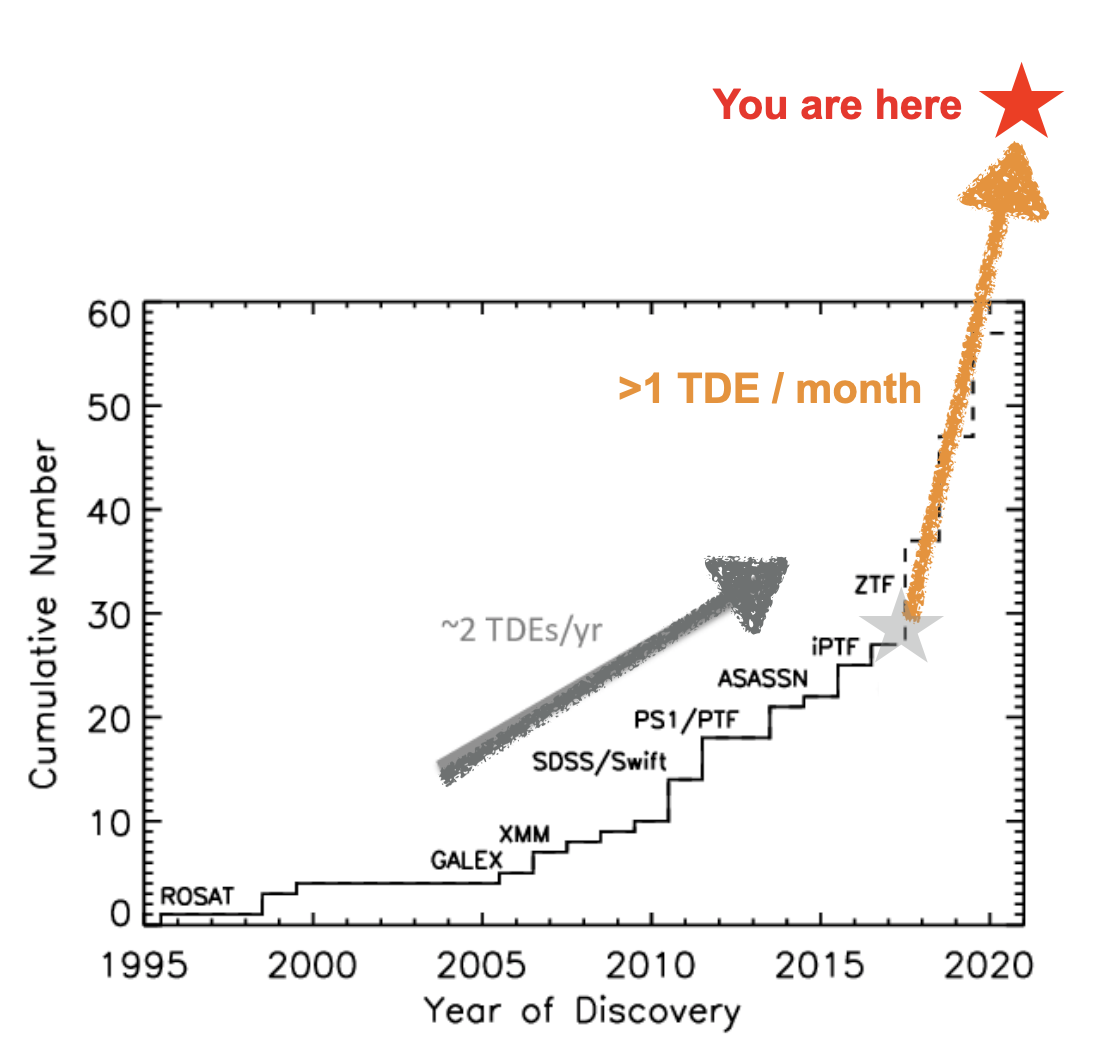}
	\caption{Prediction of TDE discoveries from \cite{ztf_19_science}, adapted to reflect the subsequent detection rates.}
	\label{fig:tde_disc}
\end{figure}

Fortunately, this contribution was instead presented at the 37th ICRC in 2021, at a time when this new era of high-frequency TDE discovery had arrived. ZTF published interim results from the first 18 months of the survey in 2020, reporting the detection of 17 TDEs \cite{van_velzen_20}. ZTF alone is now finding TDEs at a rate > 1 per month, though ZTF is of course not operating alone. TDEs are also being discovered by other optical surveys including PanSTARRS \cite{gezari_12}, ASAS-SN \cite{holoien_16}, GAIA \cite{gaia_tde_18} and ATLAS \cite{nicholl_19}, X-ray surveys such as eROSITA \cite{erosita_tde_21}, as well as radio surveys such as CNSS \cite{anderson_20} and FIRST/VLASS \cite{ravi_21}. As can be seen in Figure \ref{fig:tde_disc}, the sample of confirmed and probable TDEs has vastly increased to more than 70, a doubling since the 36th ICRC. Moreover, these ZTF-detected TDEs are generally detected well before peak with high-cadence observations, have uniform multi-wavelength coverage at UV and X-ray wavelengths \cite{van_velzen_20}, as well as some radio coverage. 

Given this wealth of new data, one can consider where TDEs falls in the broader `zoo' of optical transient populations. 
%A recent work considering the potential high-energy emission of shock-powered transients highlighted that TDEs output enormous rates of energy into the universe CITE. 
While TDEs are intrinsically rare phenomena, with rates less than 1\% of the core-collapse supernova rate, each individual TDE can be extremely bright, reaching optical outputs one thousand times brighter than a typical supernova \cite{fang_20}. In aggregate, TDEs thus output enormous rates of energy into the universe. One might naturally ask whether this population also contributes higher-energy emission, namely cosmic rays and neutrinos.

\section{High-Energy Neutrinos}

The field of neutrino astronomy has undergone a similar transformation in recent years. A diffuse flux of high-energy neutrinos was first discovered by IceCube in 2013 \cite{ic_astro_13}, launching an ongoing search for sources of these astrophysical neutrinos. Much focus at the 36th ICRC was devoted to the then-recent identification of the flaring blazar TXS 0506+056 as the first probable neutrino source \cite{ic_txs_mm_18, kappes_19}. That association was enabled by the IceCube Realtime program \cite{ic_realtime_17}, in which the arrival time and direction of probable astrophysical neutrinos are automatically reported in low-latency via the Gamma-ray Coordination Network (GCN) framework \footnote{\url{https://gcn.gsfc.nasa.gov/gcn/gcn3_circulars.html}}.

IceCube also announced at the 36th ICRC that the previous streams of high-energy neutrino alerts, High-Energy Starting Event (HESE) and Extremely-High Energy (EHE)  events, had just been replaced by new unified `Gold' and `Bronze' event streams in June 2019 \cite{ic_realtime_19}. These new alerts promised to provide a substantially elevated rate of high-quality neutrino alerts. 

\begin{figure}[!ht]
	\centering \includegraphics[width=0.65\textwidth]{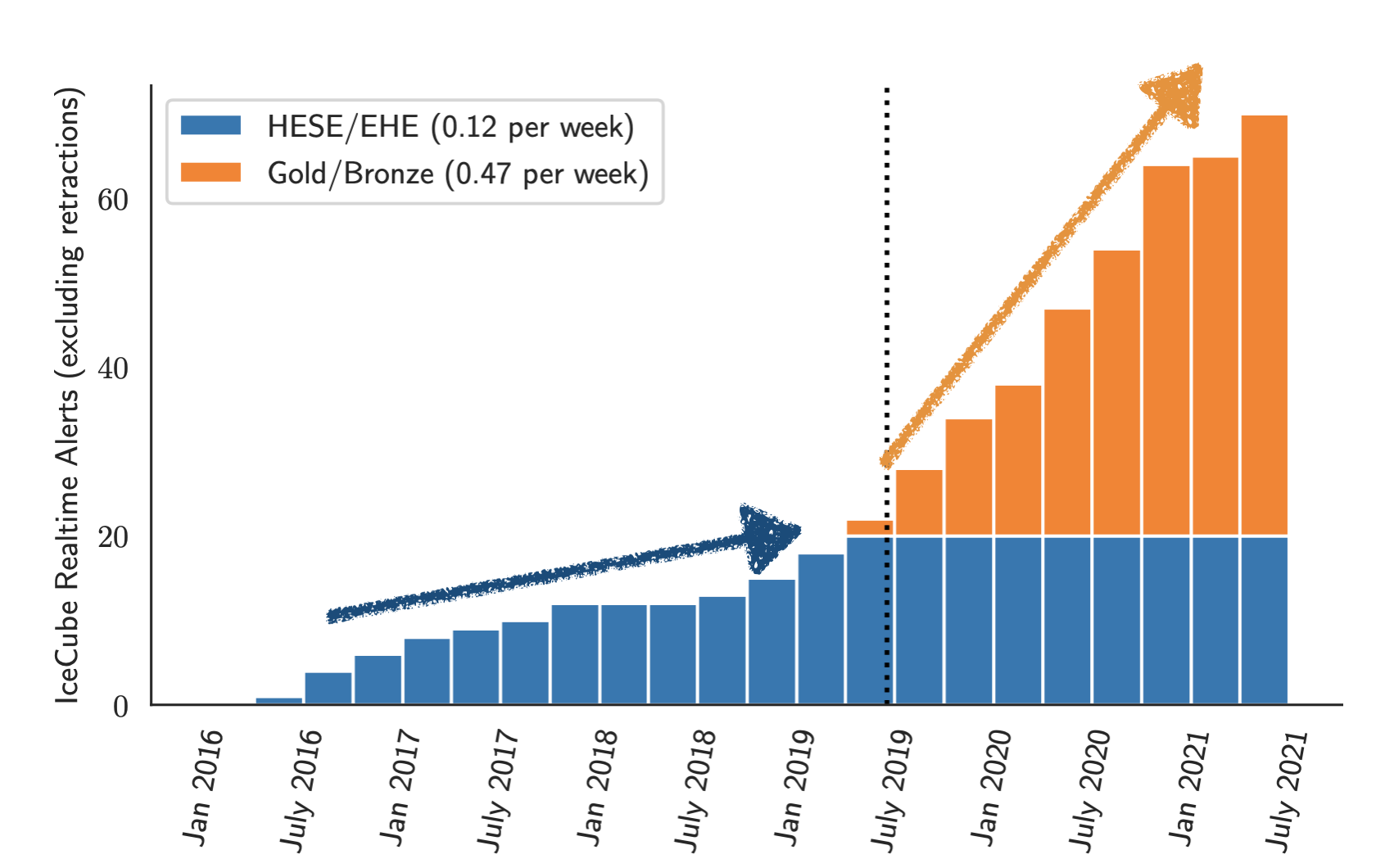}
	\caption{The cumulative number of public IceCube neutrino alerts, excluding retractions. The dotted line illustrates the transition from HESE/EHE (blue) to Gold/Bronze (orange) in June 2019.}
	\label{fig:ic_alerts}
\end{figure}

As can be seen in Figure \ref{fig:ic_alerts}, this promise has been delivered. The cumulative rate of public neutrino alerts is plotted from the start of the program in 2016 to the 37th ICRC in July 2021, excluding those alerts subsequently retracted. The alert rate has quadrupled with the transition to Gold/Bronze alerts, with the total sample size increasing more than threefold in the period since the 36th ICRC. Neutrino astronomy is thus rapidly moving to a regime of high-statistics public datasets, presenting the community with enormous opportunities to perform searches for astrophysical neutrino sources. 

\section{The Zwicky Transient Facility}

\begin{figure}[!ht]
	\centering \includegraphics[width=0.65\textwidth]{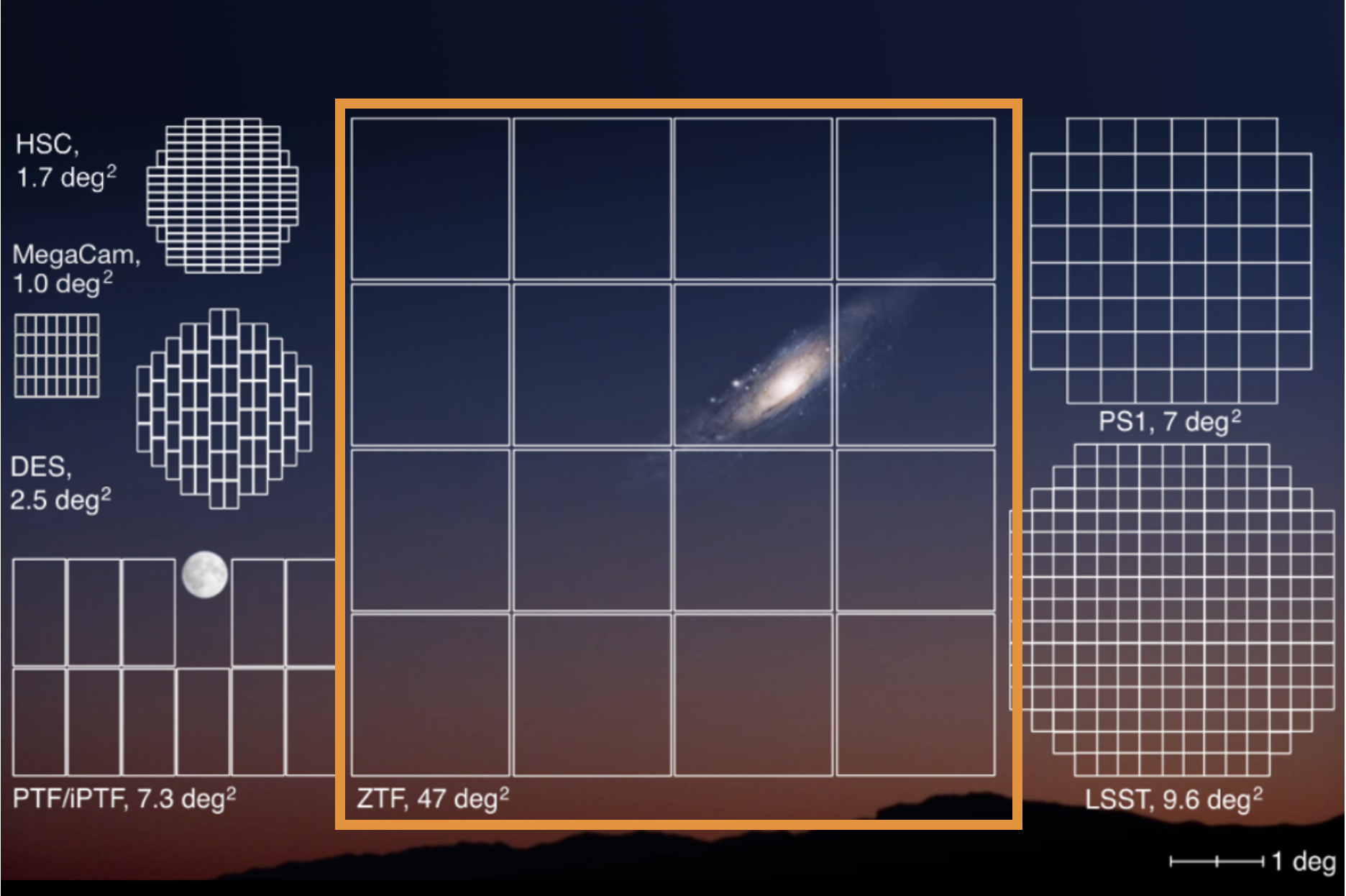}
	\caption{The ZTF field-of-view (orange), in comparison to other major telescopes \cite{laher_18}.}
	\label{fig:ztf_fov}
\end{figure}

One such search for neutrino sources is performed by ZTF, an optical telescope located on Mt. Palomar, California \cite{ztf_19_science}. ZTF is notable for its enormous 47 sq. deg. field-of-view, which dwarfs all other major optical telescopes, as can be seen in Figure \ref{fig:ztf_fov}. The ZTF design was optimised for volumetric survey speed \cite{ztf_system}, i.e to observe the largest possible universe volume in the shortest possible time. What that means in practise is that ZTF now scans the entire accessible Northern Sky every 2 nights as part of a public survey \cite{ztf_survey_19}, to a depth of 20.5 mag in g-band and r-band. ZTF provides a comprehensive accounting of the dynamic night sky, and is thus a powerful tool for the discovering of optical transients such as supernovae or TDEs.  

ZTF is also well-suited to multi-messenger searches for electromagnetic counterparts to high-energy neutrinos, a task performed with our \emph{ZTF neutrino follow-up program} \cite{bran}. IceCube neutrino alerts have a typical localisation of $\sim$10-20 sq. deg., limited by systematic uncertainties related to the optical properties of the Antarctic ice \cite{ic_resimulations_21}. Such regions are generally covered by 1 or 2 ZTF fields. We trigger dedicated Target-of-Opportunity (ToO) observations of these fields in response to a public IceCube alert. 

These observations are then processed by our multi-messenger data analysis pipeline, \emph{nuztf}, built using the Ampel software framework \cite{ampel, bran, nuztf}. Though ZTF will typically detect $\sim$500,000 objects on a given night, this number can be substantially reduced by requiring both spatial coincidence (we restrict ourselves to objects located within the 90\% localisation reported by IceCube) and temporal coincidence (we only consider objects which are detected at least once after the neutrino detection). For coincident objects, after rejecting stars and solar-system objects, we aim to spectroscopically classify all remaining candidates. These are checked to see whether they belong to populations that have been proposed as possible cosmic-ray and neutrino sources. Such populations include Gamma-Ray Bursts (GRBs) \cite{waxman_bahcall_97_grb}, supernovae with evidence of circumstellar medium interaction \cite{murase_csm_sn_11} or choked jets \cite{senno_choked_jets_16}, Active Galactic Nuclei (AGN) \cite{stecker_91} and in particular blazars \cite{mannheim_93}, and TDEs \cite{farrar_09, farrar_14, dai_17, senno_17, Biehl_tde_uhecr, guepin_18, hayasaki_19, winter_bran_21, winter_icrc_21, murase_tde_20, liu21_bran}.

\begin{figure}[!ht]
	\centering 
	\includegraphics[width=0.53\textwidth]{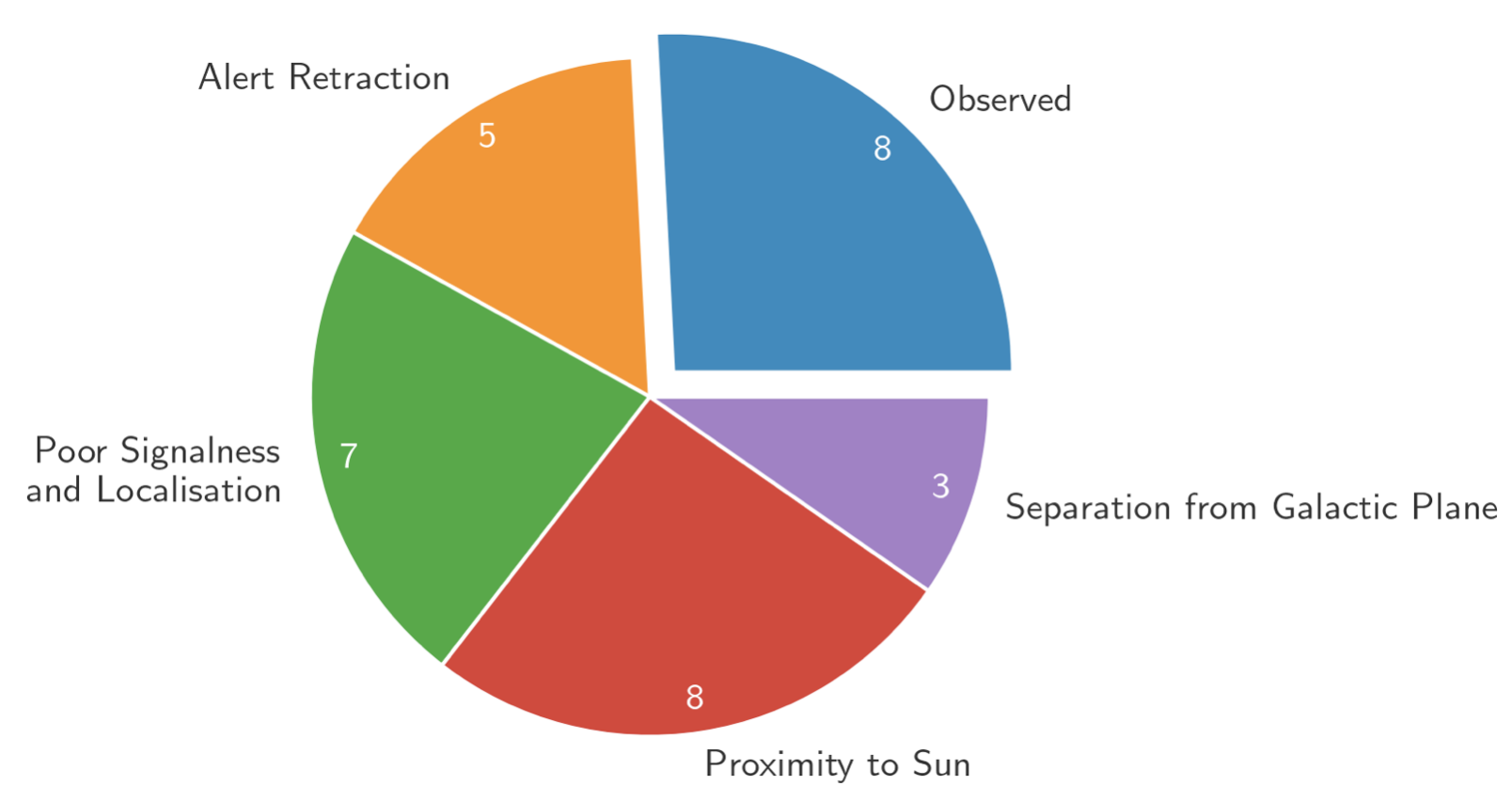}
	\includegraphics[width=0.45\textwidth]{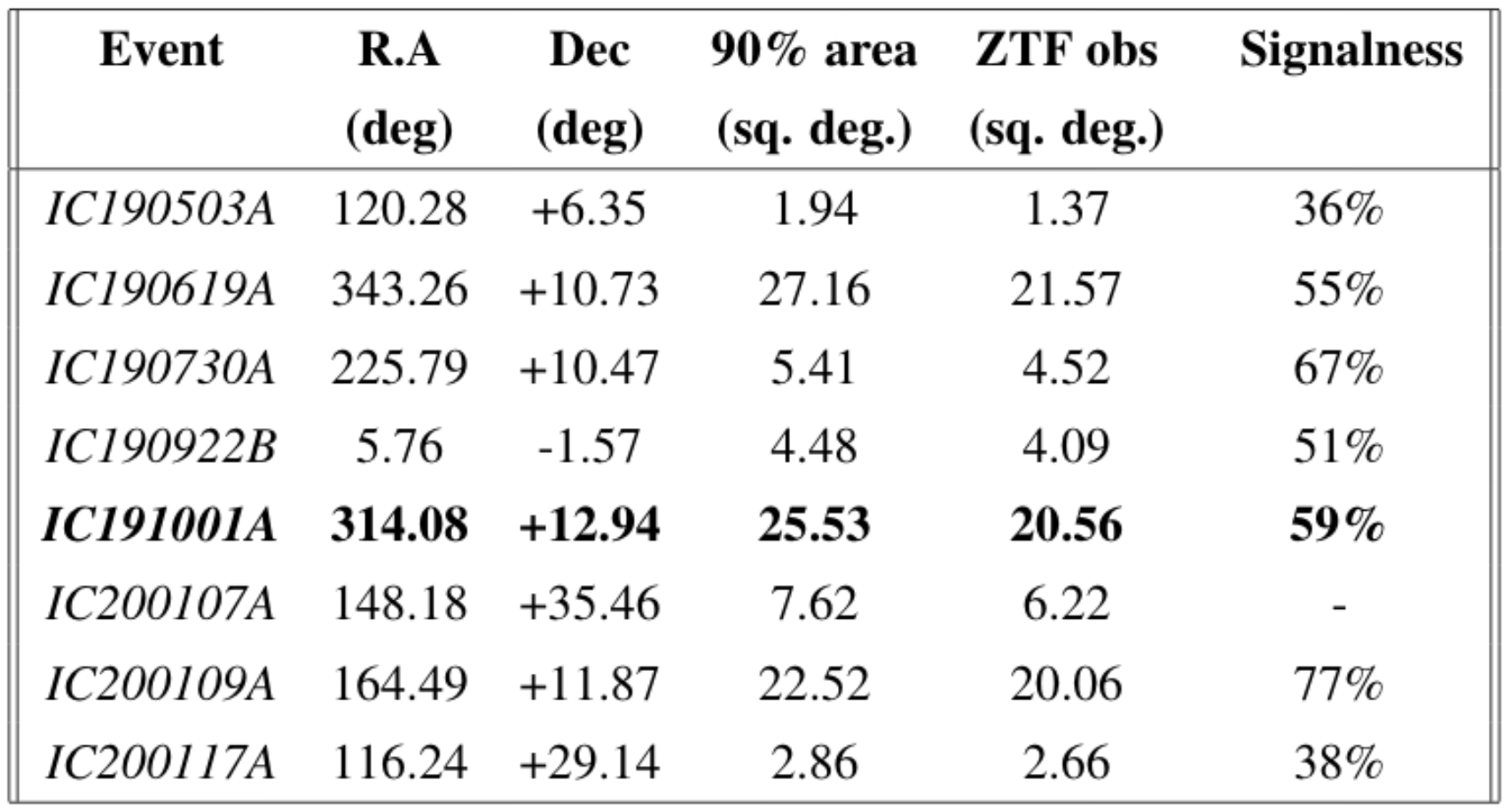}
	\caption{Left: Breakdown of ZTF observation status for IceCube neutrino triggers, as of February 2020. Right: Properties of the neutrinos observed in that period, from \cite{bran}. IC191001A is highlighted in bold. No signalness was reported for IC200107A. }
	\label{fig:ztf_stats_bran}
\end{figure}

A full breakdown of the ZTF neutrino follow-up program as of February 2020 is given in Figure \ref{fig:ztf_stats_bran}. Out of 31 alerts issued by IceCube since ZTF first light in March 2018, 8 were followed up by ZTF \cite{bran}. The remainder were either retracted, too close to the sun, or did not pass our ZTF trigger criteria. We prioritise those neutrino alerts which are well localised (90\% area less than 10 sq. deg.) or have a high probability to be astrophysical (signalness > 50\%). Of those neutrino alerts observed by ZTF, a compelling electromagnetic counterpart was only found for IC191001A \cite{bran}, highlighted in bold in Figure \ref{fig:ztf_stats_bran}. For this event, the TDE AT2019dsg was found in spatial and temporal coincidence by our program.

\section{AT2019dsg}

\begin{figure}[!ht]
	\centering
	\includegraphics[width=0.75\textwidth]{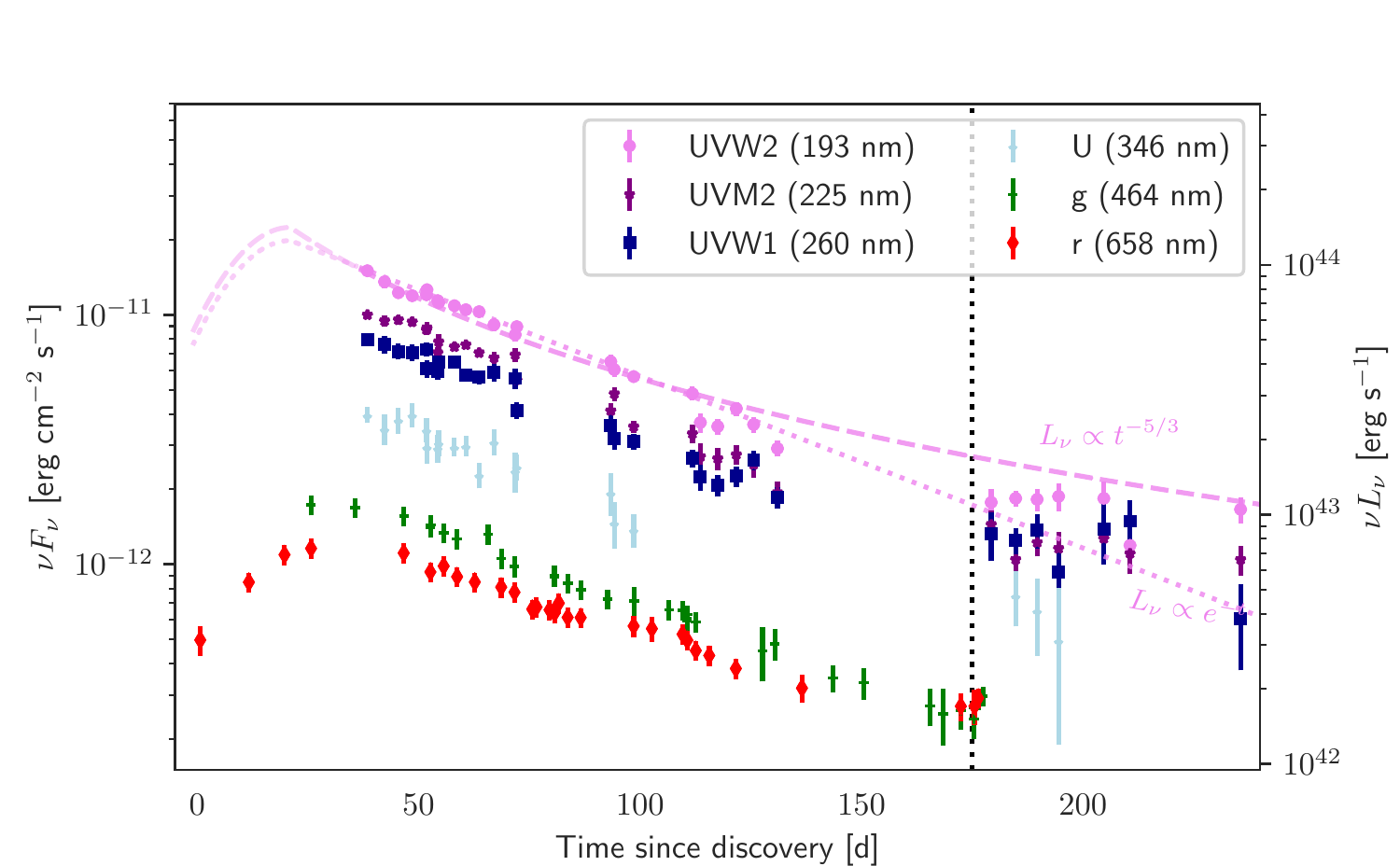}
	\caption{UV/Optical lightcurve of AT2019dsg, from \cite{bran}. The arrival of IC191001A is marked by the dashed line.}
	\label{fig:bran_uv_opt}
\end{figure}

AT2019dsg was first discovered by ZTF in April 2019, as part of a systematic search for TDEs with ZTF \cite{van_velzen_20}, and was detected continuously in the subsequent months. As for all ZTF candidate TDEs, the source was also observed repeatedly by UltraViolet/Optical Telescope (UVOT) \cite{swift_uvot_05}, on board NASA's Neil Gehrels \textit{Swift} Observatory \cite{swift_04}, providing broad coverage of the UV as well as optical emission. These observations are summarised in Figure \ref{fig:bran_uv_opt}.

The neutrino IC191001A was detected approximately 170 days after the TDE discovery \cite{bran}, as illustrated by the dashed line in Figure \ref{fig:bran_uv_opt}. Following the identification of AT2019dsg as a candidate neutrino source, additional observations were conducted with Swift-UVOT. The source continued to be detected at these later times, with an apparent plateau in emission consistent with the formation of an accretion disk \cite{bran}. More broadly, like most TDEs, the UV/optical emission of AT2019dsg was well-described by thermal emission from a blackbody of temperature $\sim10^{4.6}$K and radius $\sim10^{14.5}$cm, reaching a peak luminosity of $\sim10^{44.5}$ erg s$^{-1}$. This inferred temperature was somewhat hotter than a typical TDE \cite{van_velzen_20}. Relative to the broader ZTF TDE population, AT2019dsg was both relatively bright and also relatively close (z=0.051, corresponding to a luminosity distance of 230 Mpc), meaning that AT2019dsg had second-brightest bolometric energy flux of all ZTF TDEs \cite{van_velzen_20}. Accounting for all neutrino follow-up campaigns listed in Figure \ref{fig:ztf_stats_bran}, the probability of finding such a bright TDE by chance with our neutrino follow-up program is just 0.2\%.

\begin{figure}[!ht]
	\centering
	\includegraphics[width=0.75\textwidth]{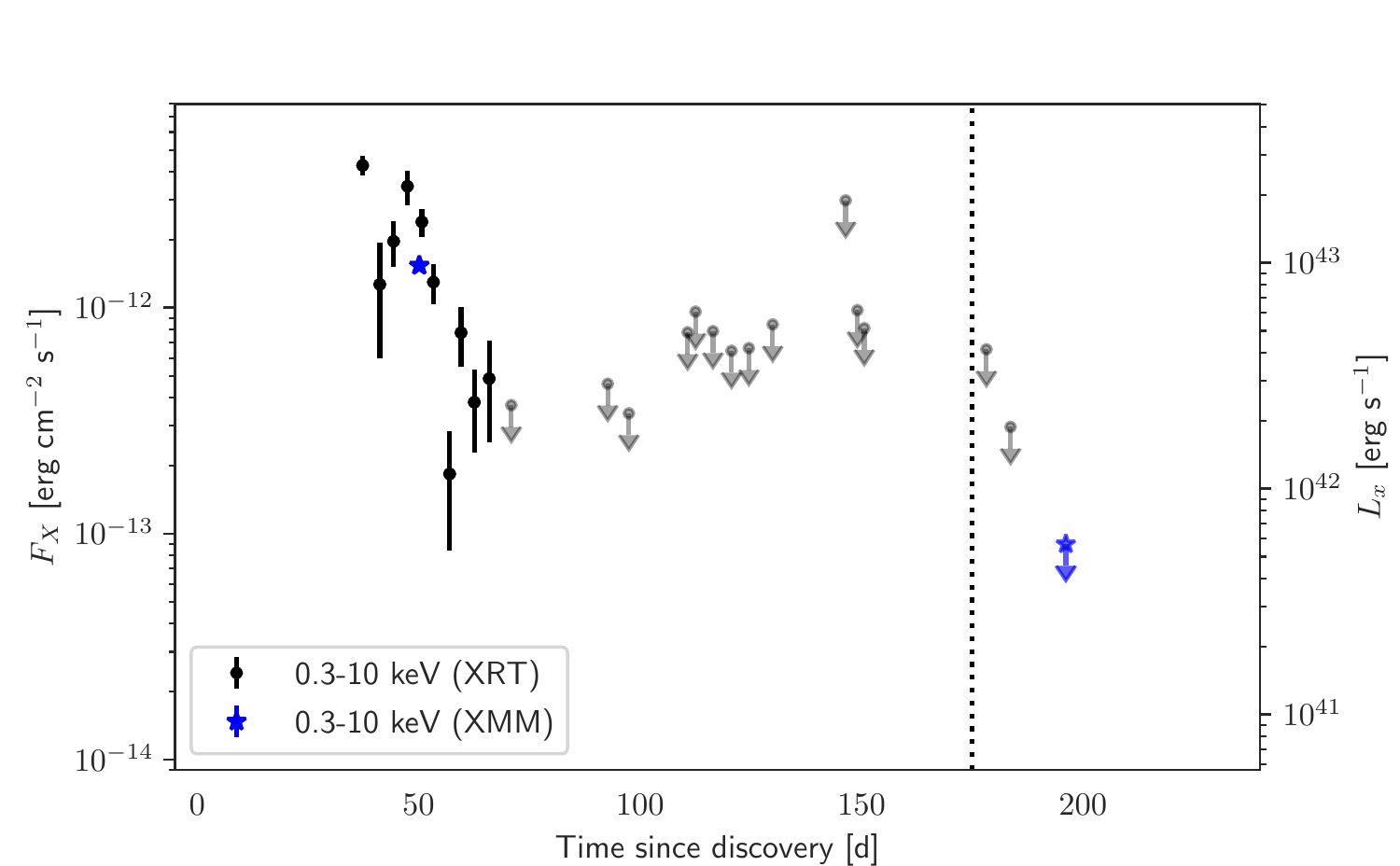}
	\caption{X-ray lightcurve of AT2019dsg, with XRT and XMM data, from \cite{bran}. The arrival of IC191001A is marked by the dashed line.}
	\label{fig:bran_xray}
\end{figure}

AT2019dsg was also detected at X-ray wavelengths by the X-Ray Telescope (XRT) \cite{swift_xrt_05}, another telescope on board the Neil Gehrels \textit{Swift} Observatory. It was further detected by the X-ray Multi-Mirror Mission (XMM)-Newton telescope \cite{xmm_01}, as well as Neutron star Interior Composition ExploreR (NICER) \cite{nicer_16,cannizzaro_21}. As can be seen in Figure \ref{fig:bran_xray}, AT2019dsg was an initially-bright source which faded extremely rapidly over the course of observations. Such a rapid fading could be explained by obscuration along the observer line of sight, or due to emission from a cooling accretion disk \cite{bran}. Such behaviour is uncommon for a TDE. Only four of the seventeen optically-selected ZTF TDEs were detected by XRT observations, and those four exhibited diverse observational properties at these wavelengths \cite{van_velzen_20}.

\begin{figure}
	\begin{subfigure}{0.45\textwidth}
		\includegraphics[width=\linewidth]{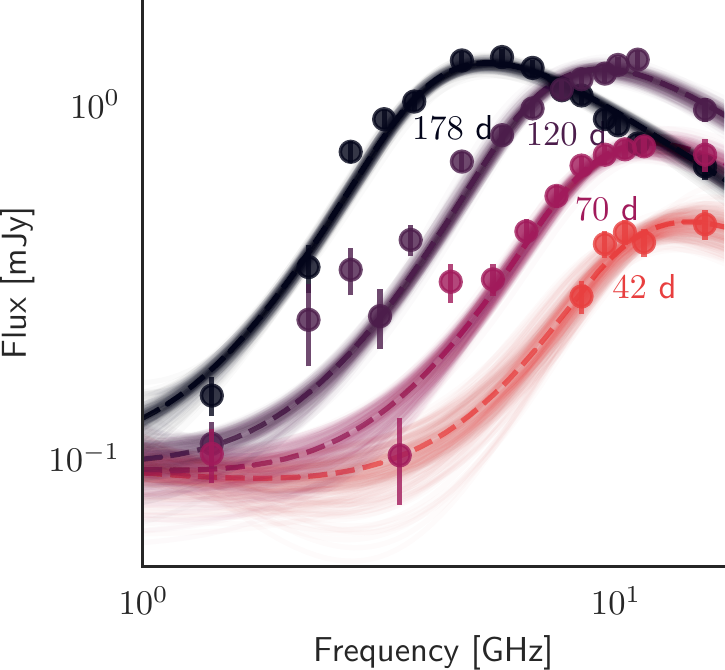}
	\end{subfigure}
	\begin{subfigure}{0.40\textwidth}
		\includegraphics[width=\linewidth]{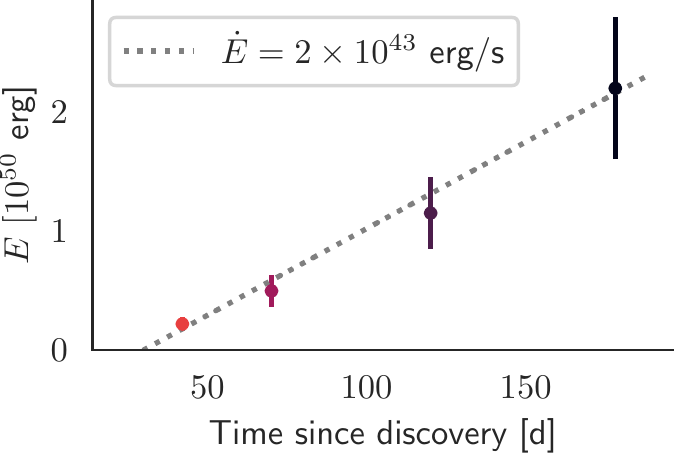}
		\includegraphics[width=\linewidth]{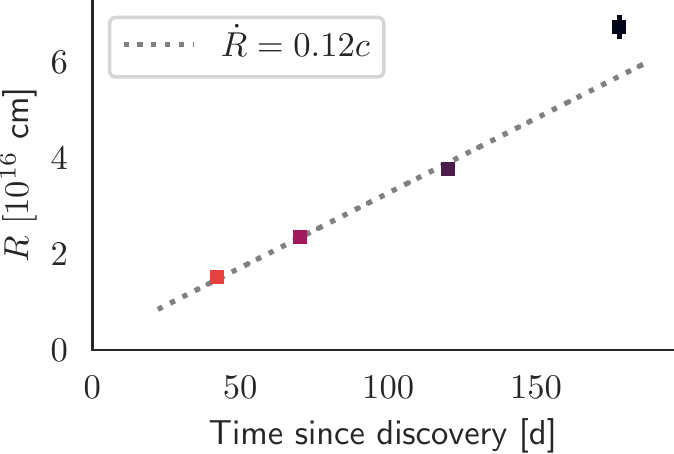}
	\end{subfigure}
	\caption{Left: Radio data of AT2019dsg, including AMI-LA, VLA and MeerKAT, from \cite{bran}. Upper Right: Inferred outflow energy. Lower Right: Inferred outflow radius.}
	\label{fig:bran_radio}
\end{figure}

Another unusual property of AT2019dsg was the accompanying radio detection, a feature shared by only a handful of other TDEs \cite{radio_tde_summary}, with emission lasting through to the time of neutrino detection \cite{bran}. Observations from Arcminute MicroKelvin Imager (AMI)-LA \cite{ami_08, ami_18}, MeerKAT \cite{meerkat_16} and Karl G. Jansky Very Large Array (VLA) \cite{vla_80} are summarised in Figure \ref{fig:bran_radio}. These observations have been interpreted as synchrotron emission from a mildly-relativistic outflow \cite{bran, cendes_21, mohan_21, matsumoto_21}. Others have interpreted this as possible emission from a relativistic jet \cite{liu21_bran, winter_bran_21}, though this has been disfavoured by others \cite{mohan_21, cendes_21}. In any case, the radio data confirm long-lived non-thermal emission in AT2019dsg \cite{bran}, and provide a conservative baseline for non-thermal energy output of the source. The extended duration of the non-thermal emission may also explain the relatively late neutrino detection. Whether such outflows are common or ubiquitous in TDEs remains an open question \cite{radio_tde_summary}. 

\section{Neutrinos from Tidal Disruption Events}

\begin{figure}[!ht]
	\centering
	\includegraphics[width=0.75\textwidth]{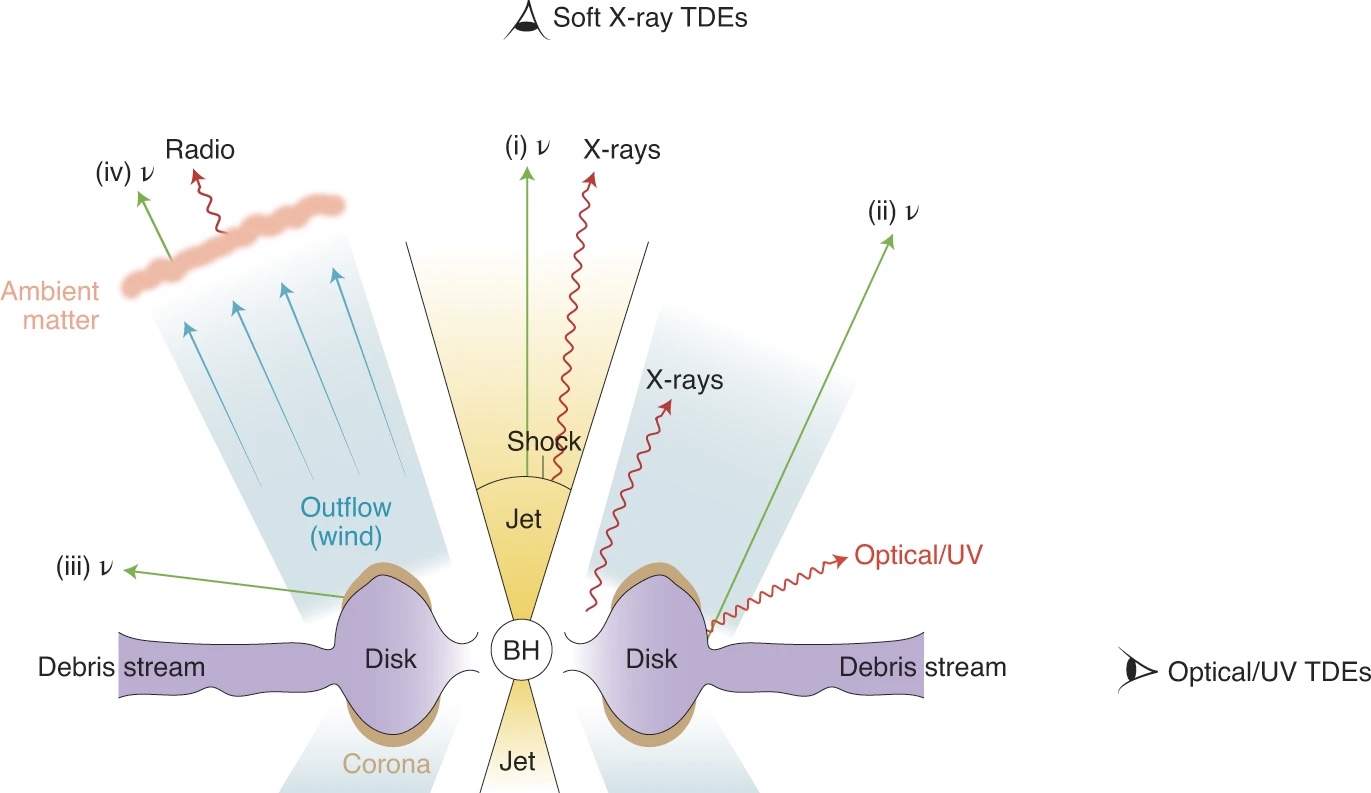}
	\caption{Schematic of possible neutrino emission zones in TDEs, from \cite{hayasaki_21}.}
	\label{fig:tde_nu}
\end{figure}

TDEs have been long-predicted to be sources of cosmic rays and neutrinos \cite{farrar_09, farrar_14, dai_17, senno_17, Biehl_tde_uhecr, guepin_18, hayasaki_19, winter_bran_21, winter_icrc_21, murase_tde_20, liu21_bran}, and were thus always a prime target of our ZTF neutrino follow-up program \cite{ztf_19_science}. However, in light of the identification of AT2019dsg as a probable neutrino source, numerous models have been updated to reflect our increased knowledge of TDEs and their observational properties. As illustrated in Figure \ref{fig:tde_nu}, these neutrino production models can be broadly grouped into four possible emission zones \cite{hayasaki_21}:

\begin{enumerate}[i]
	\item Relativistic jets \cite{winter_bran_21, winter_icrc_21, liu21_bran}
	\item The accretion disk \cite{hayasaki_19}
	\item The disk corona \cite{murase_tde_20}
	\item The wind/outflow \cite{murase_tde_20}
\end{enumerate}

In general, models agree that the conditions appear consistent with requirements for both PeV neutrino production, and the subsequent detection of a neutrino alert from such a TDE with IceCube. The observations thus suggest that the TDE population contributes to the astrophysical neutrino flux, though the ZTF observations would be compatible with a contribution as small as $2$\% of the astrophysical neutrino alerts \cite{bran}. Whether TDEs emit neutrinos at this level, or emit a larger fraction of the PeV neutrino flux, remains an open question. IceCube has separately limited the contribution of TDEs to no more than 39\% of the total, under the assumption that TDEs are neutrino standard candles following an unbroken E$^{-2.5}$ power law \cite{stein_19}. There is much open space to explore between these two constraints.

One method to further probe neutrino emission from TDEs is via direct neutrino correlation searches. Beyond IceCube, one such search with the ANTARES Neutrino Telescope did not find any evidence of excess TeV-PeV neutrino emission from AT2019dsg, though the corresponding neutrino flux predictions outlined above lay below the sensitivity of the ANTARES analysis \cite{antares_tde_21, illuminati_21}. The Baikal-GVD neutrino telescope reported preliminary indications of a possible neutrino excess from AT2019dsg, but analysis of this source is still ongoing \cite{baikail_tde_21}. 

One unresolved question is `how special is AT2019dsg?', which remains essential for understanding neutrino emission from the broader TDE population. However, as for all observations based on neutrino alerts, this is challenging to answer because of the substantial \emph{Eddington Bias} that affects all inferences of neutrino flux from individual events \cite{nora_eddington_19}. Based on one such association, it can be inferred that the ZTF TDEs cumulatively produce sufficient neutrino flux for a detection. However, it is not possible to state whether AT2019dsg is an exceptionally bright neutrino source that generates this flux alone, or merely belongs to a broader subpopulation of dim sources which in aggregate produce a detectable flux. In the latter case, the fact that AT2019dsg in particular was detected would be due to random chance, with other neutrino-emitting TDEs also having small probabilities to be detected by IceCube. In any case, the emission of neutrinos from the broader TDE population is a concrete hypothesis which can be tested and falsified.

\section{The ongoing search for neutrino sources}

\begin{figure}[!ht]
	\centering
	\includegraphics[width=0.65\textwidth]{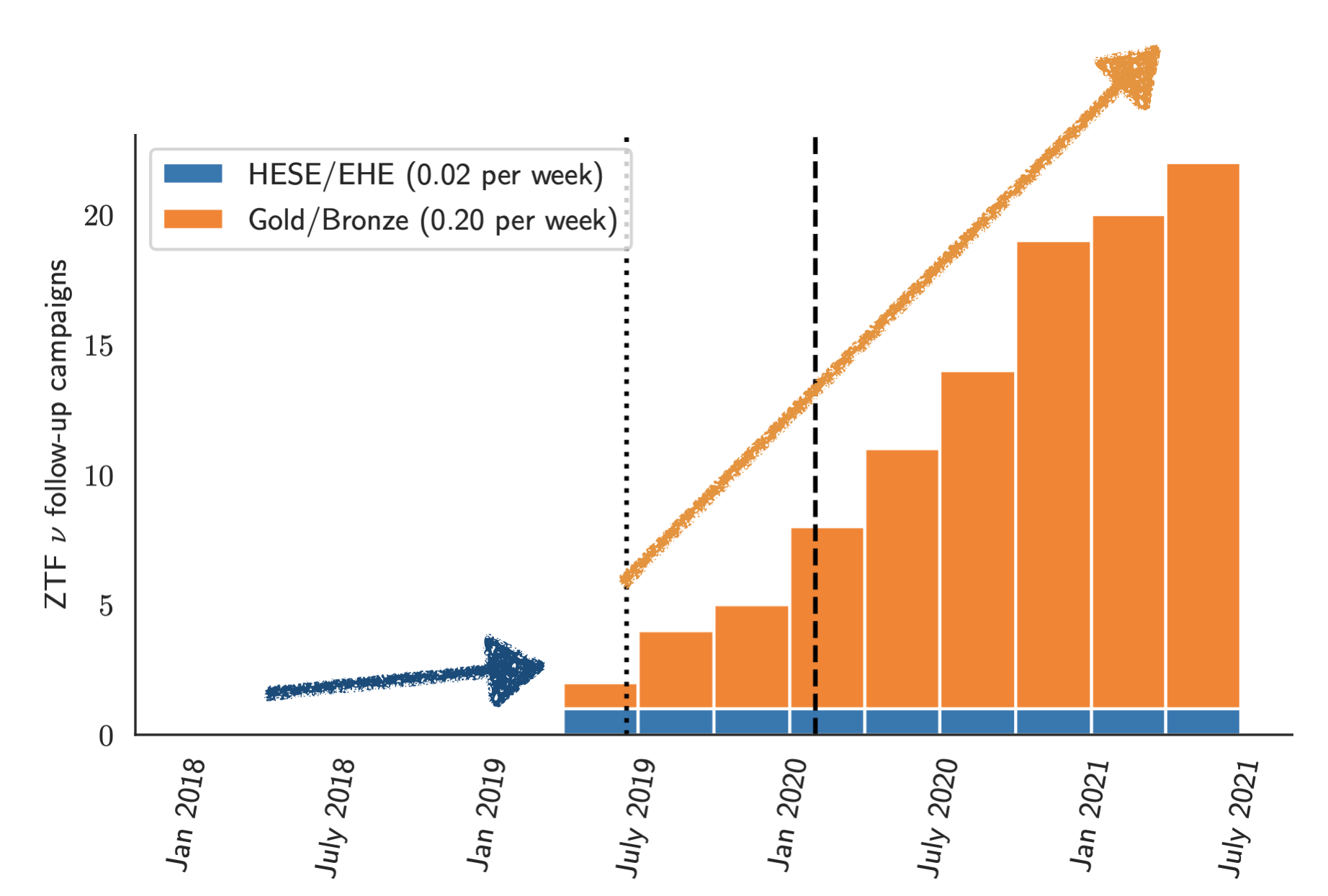}
	\caption{ZTF neutrino follow-up campaigns, as a function of time. The dotted line illustrates the transition from HESE/EHE (blue) to Gold/Bronze (orange) in June 2019. The dashed line indicates the end of data period analysed in Figure \ref{fig:ztf_stats_bran}.}
	\label{fig:ztf_nu_rate}
\end{figure}

Further observations should resolve the questions related to the TDE emission scenarios. Observations have continued at a substantially elevated rate following the commencement of IceCube Gold/Bronze alerts, which also have a more favourable hemispheric distribution for ZTF accessibility and a lower retraction rate. As can be seen in Figure \ref{fig:ztf_nu_rate}, ZTF now averages one new neutrino follow-up campaign every $\sim$5 weeks. This corresponds to more than 40\% of all IceCube neutrino alerts. In particular, in contrast to the 8 neutrinos illustrated in Figure \ref{fig:ztf_stats_bran}, 22 campaigns have now been completed as of June 2021.

\begin{figure}[!ht]
	\centering
	\includegraphics[width=0.65\textwidth]{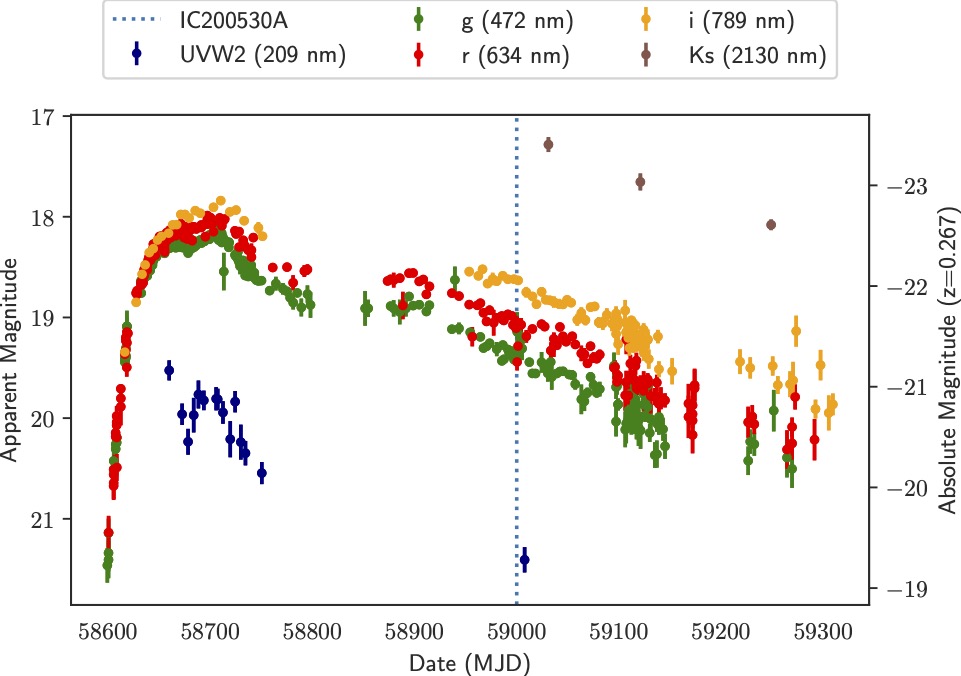}
	\caption{UV/Optical/IR lightcurve of AT2019fdr. The arrival of IC200530A is marked by the dashed line.}
	\label{fig:tywin_lightcurve}
\end{figure}

In the course of those additional ZTF campaigns, a second candidate neutrino TDE has since been identified \cite{ic200530a_ztf}. This object, AT2019fdr, was found by ZTF coincident with high-energy neutrino IC200530A. As can be seen in Figure \ref{fig:tywin_lightcurve}, AT2019fdr is a bright long-duration flare first detected by ZTF one year prior to the neutrino arrival time. AT2019fdr was classified as a probable TDE based on the basis of observed spectroscopic and photometric properties, as part of a systematic study of long-duration flares in Narrow-Line Seyfert 1 galaxies with ZTF \cite{frederick_20}. However, an extreme AGN flare origin for AT2019fdr cannot be excluded.

\begin{figure}[!ht]
	\centering
	\includegraphics[width=0.5\textwidth]{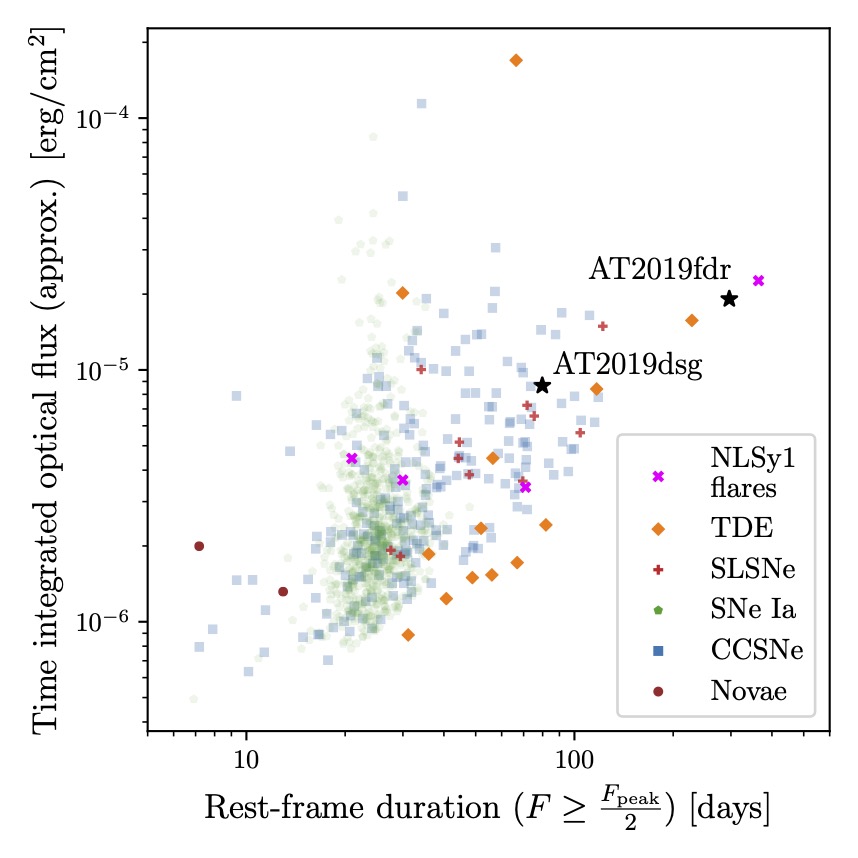}
	\caption{Time-integrated flux and rest frame duration of AT2019fdr and AT2019dsg, in comparison to the broader sample of transients detected by ZTF. }
	\label{fig:ztf_transients}
\end{figure}

The detection of a single neutrino-TDE coincidence with our ZTF program due to random chance cannot be excluded, though it remains unlikely. The detection of a second neutrino-TDE coincidence reduces this probability even further. The improbability of two random coincidence can be seen even more clearly when AT2019dsg and AT2019fdr are compared to the broader sample of transients detected by ZTF, as shown in Figure \ref{fig:ztf_transients}. Relative to transients detected by the unbiased ZTF Bright Transient Survey \cite{ztf_bts_1, ztf_bts_2}, both AT2019dsg and AT2019fdr have atypically bright time-integrated optical fluxes. Indeed, they are outliers even relative to the populations of TDEs and flares in Narrow-Line Seyfert 1 galaxies, which are themselves more intrinsically luminous than supernovae \cite{van_velzen_20, frederick_20}. These observations provide strong indications of an emerging trend, suggesting that some portion of the neutrino flux is indeed emitted by bright TDEs.

\section{Summary}

The search for sources of astrophysical neutrinos is still ongoing, and many open questions remain unanswered. Dedicated neutrino follow-up programs like the one operated by ZTF can be powerful tools to address these questions. The ZTF program has already identified the bright TDE AT2019dsg as a probable neutrino source. Conditions in this TDE appear consistent with production of a $\sim$0.2 PeV neutrino. TDEs like AT2019dsg are being discovered in ever-increasing numbers, offering the opportunity for understanding how this source fits into the broader TDE population. In parallel, the ZTF neutrino follow-up program continues to operate, and has already identified a second candidate neutrino-TDE. With luck, the 38th ICRC may feature yet more candidate neutrino sources, and further insight into the relative contributions of different populations.

\acknowledgments{I would like to thank Anna Franckowiak, Marek Kowalski, Simeon Reusch and Sjoert van Velzen for fruitful discussion. I would also like to thank the IceCube Collaboration for publishing high-energy neutrino alerts, without which such research would not be possible. This work was supported by the Initiative and Networking Fund of the Helmholtz Association, Deutsches Elektronen Synchrotron (DESY).}

\bibliographystyle{JHEP}
\bibliography{skeleton}

%\begin{thebibliography}{99}
%\bibitem{...}
%....
%
%\end{thebibliography}

\end{document}